\documentclass[pra, superscriptaddress, twocolumn, showpacs, showkeys, preprintnumbers]{revtex4}

\usepackage{amsmath}                     %
\usepackage{amssymb}                     %
\usepackage{graphicx}                    
\usepackage{psfrag}                      
\usepackage{bm}                          

\def\abs#1{\ensuremath{\left \vert #1 \right \vert}}

\begin{document}



\title[Thermodynamically activated vortex-dipoles in 2D BEC]%
      {Thermodynamically activated vortex-dipole formation in a two-dimensional Bose-Einstein condensate}

\author{D{\'a}niel Schumayer} \email{dschumayer@physics.otago.ac.nz}
\affiliation{The Jack Dodd Centre for Photonics and Ultra-Cold Atoms, 
             Department of Physics, University of Otago, P. O. Box 56, 
             Dunedin, New Zealand}

\author{David A. W. Hutchinson}
\affiliation{The Jack Dodd Centre for Photonics and Ultra-Cold Atoms, 
             Department of Physics, University of Otago, P. O. Box 56, 
             Dunedin, New Zealand}
\affiliation{Laboratoire Kastler Brossel, {\'E}cole Normale Sup{\'e}rieure, 
             24 rue Lhomond, 75231 Paris Cedex 05, France}

\date{\today}

\begin{abstract} 

   Three distinct types of behaviour have recently been identified in
   the two-dimensional trapped bosonic gas, namely; a phase coherent
   Bose-Einstein condensate (BEC), a
   Berezinskii-{\-}Kosterlitz-{\-}Thouless-{\-}type (BKT) superfluid
   and normal gas phases in order of increasing temperature. In the
   BKT phase the system favours the formation of vortex-antivortex
   pairs, since the free energy is lowered by this topological defect.
   We provide a simple estimate of the free energy of a dilute Bose
   gas with and without such vortex dipole excitations and show how
   this varies with particle number and temperature. In this way we
   can estimate the temperature for cross-over from the coherent BEC
   to the (only) locally ordered BKT-like phase by identifying when
   vortex dipole excitations proliferate. Our results are in
   qualitative agreement with recent, numerically intensive, classical
   field simulations.

\end{abstract}

%
%
\maketitle


    Since the experimental realisation of Bose-Einstein condensation
    (BEC) \cite{Anderson1995}, tremendous advances have been made in
    exploring the physics of ultracold atomic and molecular gases. It
    is now possible to simultaneously prepare multiple condensates and
    collide them \cite{Legere1998, Thomas2004} or even slice a
    condensate into numerous pieces using optical standing-waves
    \cite{Stock2005}. Such an optical potential, in addition to the
    normal harmonic confinement, allows one to create a highly
    anisotropic trap and in this way it is possible to enter the quasi
    two- \cite{Gorlitz2001, Rychtarik2004, Smith2005, Stock2005} or
    one-dimensional \cite{Gorlitz2001} regimes.

    In lower dimensions the properties of quantum systems are
    cardinally altered. Unlike in the three-dimensional (3D) case, the
    different density of states means condensation cannot occur either
    in two- (2D) or one-dimensional (1D) homogeneous ideal Bose gases
    in the thermodynamic limit at finite temperature
    \cite{Mermin1966a, Hohenberg1967a}. 

    The application, however, of a trapping potential introduces a
    low-momentum cut off which suppresses long-wavelength fluctuations
    and allows the possibility of BEC \cite{Bagnato1991, Ketterle1996}
    below a certain critical temperature, which for the ideal 2D case
    we denote $T^{\mathrm{2D}}_{0}$.

    In the homogeneous 2D system, although BEC is absent at finite
    temperature, superfluidity can exist with algebraically decaying
    order \cite{Berezinskii1971a, Berezinskii1972a, Kosterlitz1973} in
    an interacting but spatially uniform Bose gas. This is the
    Berezinskii-{\-}Kosterlitz-{\-}Thouless (BKT) phase in which the
    gas has local order but no global phase coherence, with no uniform
    order parameter. A local order parameter can be defined and is
    known as the quasi-condensate \cite{Popov2002}.

    A similar scenario persists in the presence of external
    confinement. Three distinct regions can be identified in a
    two-dimensional trapped interacting bosonic system. At
    temperatures much below $T^{\mathrm{2D}}_{0}$ a phase coherent BEC
    appears to exist \cite{Petrov2000, Gies2004a}. At higher
    temperatures the coherence length decreases to length scales
    smaller than the system size \cite{Gies2004b} and bound 
    vortex-antivortex pairs (vortex-dipoles) begin to form
    \cite{Simula2005a} indicating a BKT-like phase. Eventually, around
    $T^{\mathrm{2D}}_{0}$, the vortex-dipoles unbind and the
    correlation length decays exponentially. This is the analogue of
    the BKT phase transition to the thermal gas \cite{Simula2006a,
    Prokofev2001}.

    A phase diagram and approximate critical temperature for the BEC
    to BKT cross-over were derived numerically in \cite{Simula2005a}
    assuming that the vortex-dipole is always at the centre of the
    trap, i.e. placed at the highest density. Here, we use simple
    analytic arguments, without this restriction allowing the
    vortex-dipole to be placed anywhere inside the Thomas-Fermi
    radius, to obtain a very simple picture of the vortex dipole
    proliferation in the gas. Our approach obviates the need for any
    detailed numerical calculations and qualitatively agrees with the
    more sophisticated approaches, including the full classical field
    simulations \cite{Simula2006a}.


    In our model, we minimize the Helmholtz free-energy, $F = U - TS$,
    where $U$ denotes the total internal energy and $S$ the entropy of
    the state, and the system has constant volume $V$ and  an absolute
    temperature $T$. We partition the gas into two subsystems, namely,
    a condensate and a thermal part, each with its own internal energy
    and entropy contribution, ($E_{\mathrm{c}}$, $E_{\mathrm{t}}$) and
    ($S_{\mathrm{c}}$ and $S_{\mathrm{t}}$), respectively. In the
    condensate part we allow for a vortex-dipole excitation, i.e. a
    bound vortex and antivortex pair. It suffices to consider only the
    effect of one pair in the condensate with the additional
    configurational entropy of the vortex dipole denoted as $\Delta
    S_{\mathrm{v}}$.

    The important contributions come from the condensate, with the
    thermal component only producing global energy shifts.
    Accordingly, $E_{\mathrm{t}}$ and $S_{\mathrm{t}}$ are only
    roughly approximated. The free energy of the system can now be
    written as $F=U-TS=(E_{\mathrm{c}}+E_{\mathrm{t}}) - T
    (S_{\mathrm{c}} +  \Delta S_{\mathrm{v}} + S_{\mathrm{t}})$.

    The model system consists of $N$ interacting bosonic particles of
    mass $m$ confined by an external non-rotating harmonic potential
    characterized by the $\lbrace \omega_{\perp}, \omega_{\mathrm{z}}
    \rbrace$ trap frequencies. The actual values of the physical
    parameters we take correspond to the experiments of
    \cite{Stock2005, Smith2005}. In both experiments the radial and
    axial trapping frequencies differ by several orders of magnitude,
    suggesting that we use a quasi-2D formalism. This assumption is
    confirmed by comparing the characteristic oscillator length and
    the s-wave scattering length \cite{Petrov2000}. Consequently, we
    assume that the wavefunction for the condensate can be factorized
    into two parts: $\Psi = \psi_{\perp}(x,y) \psi_{z}(z)$, where
    $\psi_{\perp}$ describes the state in the $xy$-plane, while
    $\psi_{z}$ is simply the ground state wavefunction of a harmonic
    oscillator in a trap characterized by $\omega_{z}$. The
    wavefunction $\psi_{\perp}$ we normalize to the condensate
    particle number $N_{0}$ and $\psi_{z}$ to unity.

    Moreover, we introduce an arbitrary spatial unit $x_{\mathrm{s}}$,
    rescaling the spatial variables and the wavefunction
    $\psi_{\perp}$ to be dimensionless. In the numerical calculation
    the spatial unit $x_{\mathrm{s}}$ was taken to be equal to the
    oscillator length $a_{{\mathrm{osc}},\perp}$ corresponding to the
    frequency $\omega_{\perp}$.

    Substituting this composite wavefunction into the usual
    Gross-Pitaevskii energy functional and evaluating all integrals
    with respect to the $z$ direction, we obtain the following
    expression for the energy $E_{\mathrm{c}}$ of the condensate
    \begin{widetext} 
       \begin{equation} 
          E_{\mathrm{c}} = N_{0} \,\,\frac{\hbar \omega_{z}}{2} +
                           \frac{\hbar^{2}}{2m x^{2}_{s}} 
                           \int{\left \lbrack 
                                \abs{\nabla^{\mathrm{2D}} 
                                \psi_{\perp}}^{2}               +
                                V_{\mathrm{ext}}^{\mathrm{2D}}
                                  ({\mathbf{r}})
                                \abs{\psi_{\perp}}^{2}          + 
                                2\sqrt{2\pi} \, 
                                \left ( 
                                   \frac{a_{\mathrm{s}}}%
                                        {a_{{\mathrm{osc}},z}}
                                \right ) 
                                \abs{\psi_{\perp}}^{4}
                                \right \rbrack d{\mathbf{r}} 
                               }, 
       \end{equation} 
    \end{widetext}	
    where $\nabla^{\mathrm{2D}} = (\partial_{x}, \partial_{y})$, and
    the potential term $V_{\mathrm{ext}}^{\mathrm{2D}}$ is
    cylindrically symmetric, $(x_{\mathrm{s}}/ a_{{\mathrm{osc}},\perp})
    (x^{2} + y^{2})$. Interaction between particles is taken into
    account via the 3D $s$-wave scattering length, $a_s$.

    The internal energy of the thermal part was approximated by the
    energy of the trapped but non-interacting Bose-gas
    \cite{Dalfovo1999}
    \begin{equation} \label{eq:EnergyOfThermalCloud}
       E_{\mathrm{t}} = (N-N_{0}) k_{\mathrm{B}} T^{\mathrm{2D}}_{0}
                        \left ( 
                            \frac{3 \zeta(4)}{\zeta(3)} 
                        \right ) 
                        \left ( 
                           \frac{T}{T^{\mathrm{2D}}_{0}} 
                        \right )^{4}
                      .
    \end{equation}
    The corresponding entropy contribution $S_{\mathrm{t}}$ can be
    derived from Eq.~\eqref{eq:EnergyOfThermalCloud} using the
    standard formulae of statistical physics. The validity of this
    approximation is also justified by experiment
    \cite{Stock2005}, where the thermal cloud was not fully within the
    2D regime. The total number of particles was close to the
    theoretical threshold \cite{Fernandez2002}, above which the atoms
    squeeze out to the third dimension even if the condensate
    dimension in that direction is smaller than the healing length. 

    Naturally, a more sophisticated treatment is possible and in
    particular, the presence of the thermal cloud could significantly
    effect the energy of the vortex core \cite{Rajagopala2004}. To go
    beyond our simple treatment would require a numerically intensive,
    self-consistent, finite-temperature Hartree-Fock-Bogoliubov
    treatment \cite{Griffin1996, Hutchinson1997, Simula2002} which is
    not in the spirit of this exercise. Here we aim at the simplest
    possible estimate of the critical temperature for vortex dipole
    proliferation and we do not expect inaccuracies in the vortex core
    energy at high temperatures to qualitatively effect our
    conclusions. This expectation is confirmed by comparison with the
    (numerical) classical field simulations of Simula and
    Blakie \cite{Simula2006a, Hutchinson2006}.

    In order to evaluate the free energy functional $F$, we must also
    approximate the entropy, which
    \begin{figure}[b!]
       \includegraphics[width=60mm]{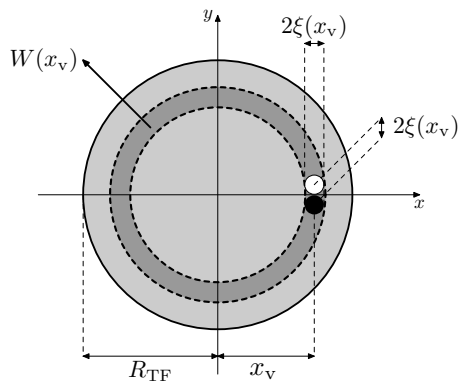}
       \caption{\label{fig:AreasForEntropy}%
                The dark ring represents the equivalent microstates
                for the vortex pair displaced at distance
                $x_{\mathrm{v}}$ from the centre.
               }
    \end{figure}
    is proportional to the number of physically available microstates.
    Only the ground state is available for particles in the
    condensate, therefore $S_{\mathrm{c}} = 0$. On the other hand, if
    a vortex-dipole is present in the condensate then its entropy can be
    estimated via a simple geometrical approach (see
    Fig.~\ref{fig:AreasForEntropy}) such that $ \Delta S_{\mathrm{v}}
    = k_{\mathrm{B}} \ln{ \! ( 2 x_{\mathrm{v}}/ \xi (x_{\mathrm{v}})
    )}$, where $x_{\mathrm{v}}$ denotes the distance of the
    vortex-dipole from the centre. The healing length characterizes
    the length over which the density depression vanishes and the wave
    function approaches its bulk value; it is defined by
    $\xi^{-1/2}(x_{\mathrm{v}}) = 4 \sqrt{2\pi} ( a_{\mathrm{s}} /
    a_{{\mathrm{osc}},z} ) \abs{\phi_{\mathrm{TF}}
    (x_{\mathrm{v}})}^{2}$ .

    It is now clear that only, $E_{\mathrm{c}}$ and $\Delta
    S_{\mathrm{v}}$, have spatial dependence and the remaining energy
    and entropy expressions simply shift the free-energy by a constant
    value at a given temperature.

    Finally, to evaluate $E_{\mathrm{c}}$ in mean-field theory, one
    needs a wavefunction for the condensate containing a vortex-dipole
    excitation. The square modulus of the wavefunction $\psi_{\perp}$
    without any excitation is chosen to have the Thomas-Fermi density
    profile on which we superpose the vortex-dipole excitation in the
    form $\psi_{\perp} ({\mathbf{r}}) =
    \phi_{\mathrm{TF}}({\mathbf{r}})  \varphi_{-1}({\mathbf{r}})
    \varphi_{1}({\mathbf{r}})$, where the vortex or antivortex
    wavefunction is
    \begin{equation*} 
       \varphi_{q}({\mathbf{r}}) = 
       \tanh{
             \!\left ( 
                  \frac{\abs{{\mathbf{r}} - {\mathbf{r}}_{q}}}%
                       {\xi({\mathbf{r}}_{q})} 
               \right)
            }
       e^{i q \Theta({\mathbf{r}}, {\mathbf{r}}_{q})} 
       \qquad 
       q = -1, +1.
    \end{equation*}
    The two values of $q$ correspond to a vortex and anti\-vortex and
    ${\mathbf{r}}_{q} = (x_{q}, y_{q})$ denotes the positions of the
    localized excitations. The hyperbolic tangent ensures the proper
    asymptotes both in the vortex core and far from the vortex. The
    phase $\Theta({\mathbf{r}}, {\mathbf{r}}_{q})$ measures the angle
    of ${\mathbf{r}}$ with respect to the $x$-axis. It gives a $2\pi$
    phase difference if one integrates along a path encircling the
    vortex core. We note here that the thermal cloud could infiltrate
    into the cores of vortex-dipoles and may enhance the expulsion of
    the condensate \cite{Rajagopala2004} which, as discussed
    previously, our model does not take fully into account. The
    vortex-dipoles appear at the edge of the condensate however, where
    the disturbance due to the thermal cloud is expected to be small.
    We therefore expect our simple model to be qualitatively accurate.

    Three different vortex-dipole configurations were analyzed, namely
    a) $(x_{-1},y_{-1}) = (x_{1},-y_{1})$, b) $(x_{-1},0) =
    (-x_{1},0)$ and c) $(x_{-1},0) = (x_{1}-\xi(x_{1}),0)$.

\begin{figure}[th!]
       \includegraphics[width=0.490\textwidth]{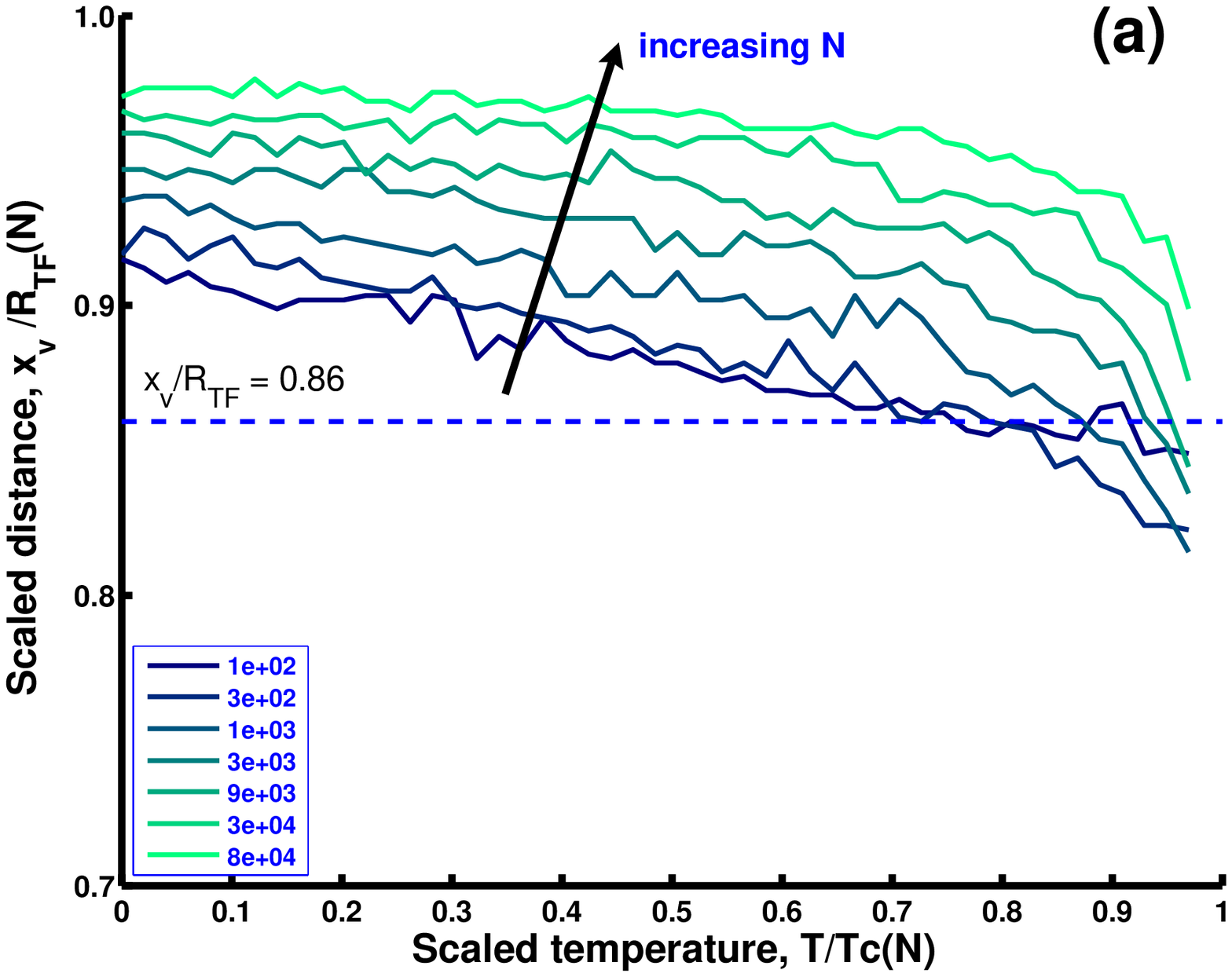}
       \includegraphics[width=0.495\textwidth]{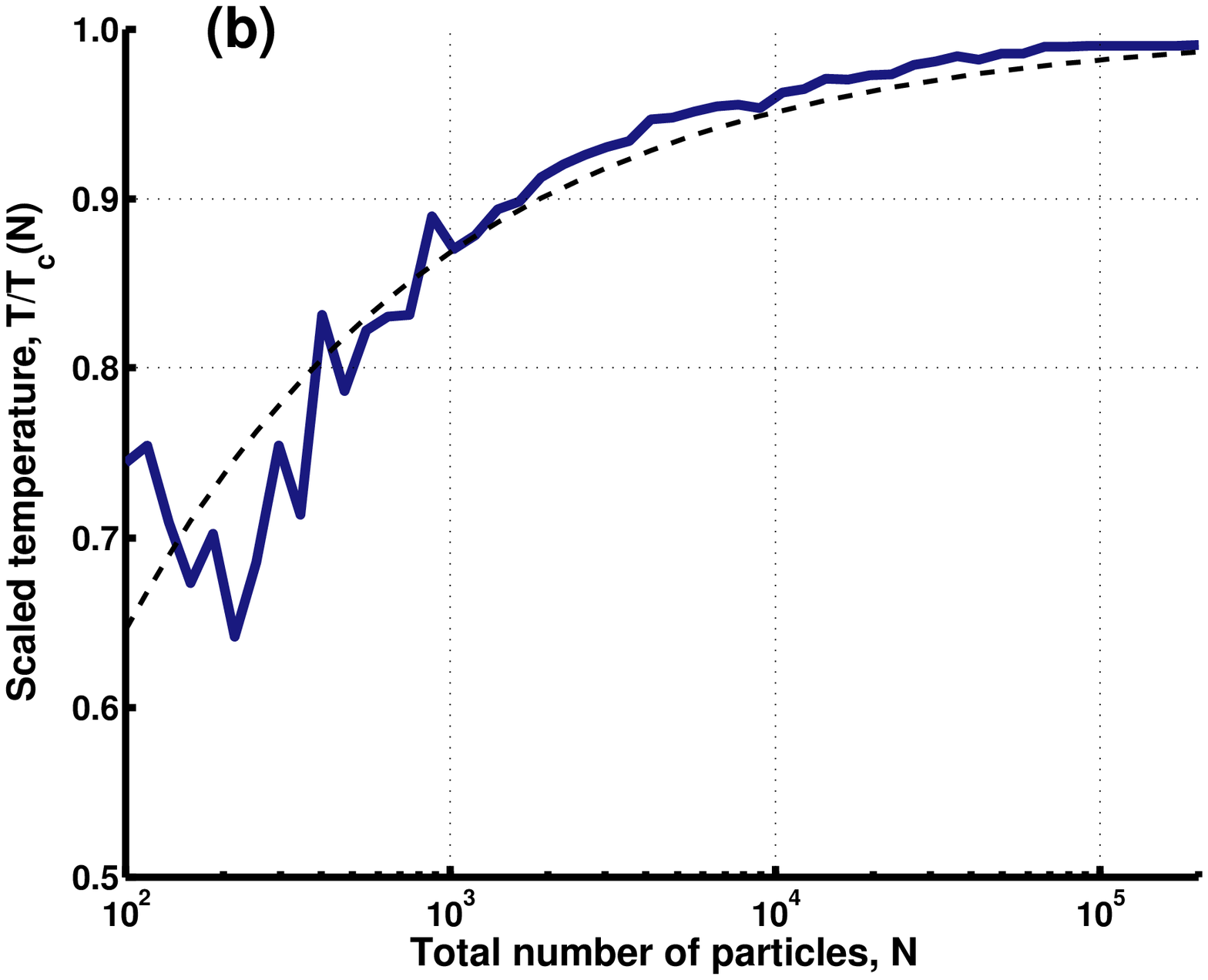} 
       \caption{\label{fig:VortexConfiguration_A_Result}
                (a) Scaled minima of free energy versus scaled 
                    temperature. Data are for configuration a).
                    The horizontal dashed line is a reference 
                    used in the text.\ %
                (b) The temperature required for a
                    vortex-dipole to penetrate into the condensate to
                    a distance of $x^{(0)}_{\mathrm{v}}$ as a function
                    of total particle number $N$. The dashed line is
                    a guide to the eye.                    
               }
    \end{figure}

    In the calculation, the total particle number $N$ and the scaled
    temperature $T/T^{\mathrm{2D}}_{0}$ were varied. For every pair of
    $N$ and $T$ we determined the position $x_{\mathrm{v}} = x_{1}$ of
    the vortex dipole to be the point where the free-energy reaches
    its minimum. These minima belonging to configuration a), are
    represented in Fig.~\ref{fig:VortexConfiguration_A_Result}(a).

    As expected, the position decreases as the temperature approaches
    the Bose-Einstein condensation temperature $T^{\mathrm{2D}}_{0}$,
    meaning that the minima of free energy are deeper inside the
    condensate. Simultaneously, however, the increasing particle
    number opposes the vortex-dipole formation well inside the
    condensate. Since vortex nucleation is not just a phase-pattern
    but also involves a density dip, more energy is required to create
    and maintain a density disturbance in a larger condensate.

    In the $T \rightarrow 0$ limit, one would anticipate that
    $x_{\mathrm{v}}$ would approach $R_{\mathrm{TF}}$. The small
    deviation from this expectation seen in
    Fig.~\ref{fig:VortexConfiguration_A_Result}(a) is attributable to
    the numerical procedure, since we require the vortex-dipole 
    remain within the Thomas-Fermi radius.

    Theoretically speaking, statistical physics allows for the
    existence of a vortex-dipole excitation inside the condensate
    phase with non-zero probability at every finite temperature, since
    the likelihood of nucleation is proportional to
    $\exp(-F/k_{\mathrm{B}}T)$. However, there exists a well-defined
    temperature $T_{\mathrm{c}}$ above which this probability becomes
    comparable with unity. It is an important question how this
    temperature varies with respect to the total number of particles.
    Using the data depicted in
    Fig.~\ref{fig:VortexConfiguration_A_Result}(a) one can infer the
    behaviour of $T_{\mathrm{c}}(N)$. Let us fix $x_{\mathrm{v}}$ at
    an arbitrary value $x^{(0)}_{\mathrm{v}}$ (in our case $0.86
    R_{\mathrm{TF}}$). Calculating the intersections of this line
    with the curves belonging to various particle numbers [see the
    horizontal dashed line in
    Fig.~\ref{fig:VortexConfiguration_A_Result}(a)], one obtains
    $(N_{i}, T_{i})$ pairs. This $T_{i}$ represents the 'critical'
    temperature for which a system with $N_{i}$ particles has a
    minimizing vortex-dipole position at the given distance
    $x^{(0)}_{\mathrm{v}}$ from the centre. If we increase the
    temperature the vortex-pair can enter further. 

    The results are represented in
    Fig.~\ref{fig:VortexConfiguration_A_Result}(b) and show that the
    temperature finally approaches the condensate critical temperature
    as the total number of particles grows. The slope of this graph
    naturally depends upon the value of
    $x^{(0)}_{\mathrm{v}}$ but the tendency itself does not and these
    findings are comparable with \cite{Simula2005a}.

    One can also conclude that it is rather difficult to observe the
    transition from BEC to BKT state experimentally. At low $N$ the
    condensate would not be dependably detectable after ballistic
    expansion, while at high $N$ the vortex-dipole formation in the
    bulk condensate phase requires a high temperature, which also
    results in a small condensate and large fluctuations.


   In conclusion, our simple model captures the physics of previous
   more complex analyses. There are two contributions to the free
   energy which depend on the position of the vortex-dipole; the
   energy of the condensate and the entropy change due to the
   excitation. One anticipates that the energy should monotonically
   decline to the value of Thomas-Fermi energy as $x_{\mathrm{v}}$
   approaches $R_{\mathrm{TF}}$, since the vortices deform the
   Thomas-Fermi density to an ever-decreasing extent. Similarly, the
   magnitude of the entropy contribution also decreases. This is
   because since the characteristic size of the vortices $\xi \propto
   n^{-1/2}$, the healing length rapidly increases. Although the area
   of the disk $W(x_{\mathrm{v}})$ of available points in which to
   place the vortices grows linearly with radius, this does not
   result in more states available for occupation due to the
   increasing size of the vortex-dipole. Consequently, these two
   competing geometrical effects are bound to lead to a minimum in
   the free energy $F$.

   A similar calculation has been performed in \cite{Zhou2004} and
   it was found that for a given angular momentum carried by the vortex
   dipole, the energy curve has a minimum inside the condensate.
   However, a direct numerical comparison with that result is not
   possible due to the differing approximations used.

   Summarizing our result we can say that
     (i)               the configurations of vortex dipoles are not
                       energetically equivalent; configuration a)
                       seems to be able to enter the bulk condensate
                       more easily than the others as the temperature
                       increases.
     (ii)              All types of configuration suggest that 
                       vortex-dipoles prefer nucleation at the 
                       edge of the condensate, and they can 
                       penetrate further in the vicinity of the
                       normal gas phase-transition ($\approx 0.8
                       -0.9 \, T^{\mathrm{2D}}_{0}$). This result
                       is consistent with and supports the possibility
                       of vortex-dipole nucleation in the experiment
                       \cite{Stock2005} at ENS, Paris and is also
                       consistent with their recent observation of
                       the cross-over from thermal gas to BKT-like
                       phase \cite{Hadzibabic2006}.
     (iii)             With increasing particle number, the
                       temperature $T_{\mathrm{c}}$ characterising
                       the vortex-dipole nucleation also becomes
                       higher and asymptotically reaches the critical
                       temperature $T^{\mathrm{2D}}_{0}$ of
                       BEC in the ideal gas. Qualitatively, the
                       same tendency was found in \cite{Simula2006a}
                       using the numerically intensive classical
                       field simulation, in which dynamical effects
                       were taken into account inherently.

   The nature of the change from BEC to BKT superfluid, i.e. whether
   this is a slow continuous change or corresponds to a sharp
   cross-over, remains an open question and is the subject of further
   investigations.


   We express our thanks for financial support from the
   Marsden Fund and from the C. N. R. S. of France.



\end{document}